\documentclass{aastex631}

\graphicspath{{./}{figs/}}

\begin{document}

\title{Resolution Dependence of Cloud-Wind Simulations}

\author[0009-0002-0079-4130]{Hannah J. Leary}
\affiliation{Department of Physics and Astronomy, University of Pittsburgh, 3941 O'Hara St, Pittsburgh, PA 15260}

\author[0000-0001-9735-7484]{Evan Schneider}
\affiliation{Department of Physics and Astronomy, University of Pittsburgh, 3941 O'Hara St, Pittsburgh, PA 15260}

\author[0000-0001-6325-9317]{Helena M. Richie}
\affiliation{Department of Physics and Astronomy, University of Pittsburgh, 3941 O'Hara St, Pittsburgh, PA 15260}

\begin{abstract}

wind-tunneand accelerationalso in bothsubsonic and ingtions run lla code, we investigate the effect that numerical resolution has on clFord evolsetup in sexplore imulWnumerical e consider upersonic and subsonic regimes. We begin with two groups of adiabatic wind tunnel simulations with wind speeds of 100 km/s and In the subsonic case, we find that a0 km/s, respectively. s per cloud radiusoisthl at five vary in the cloud evolutions:In the supersonic case the trend is more monotonic, consistent with the difference in timescales for which ram pressure acceleration dominates over mixing in the early acceleration. 

\end{abstract}

\keywords{Astrophysical fluid dynamics(101) --- Circumgalactic medium(1879) --- Galactic winds(572) --- Interstellar clouds(834)}

\section{Introduction} \label{sec:intro}

Galactic winds play a crucial role in galaxy formation, as they regulate star formation and enrich the circumgalactic medium (CGM) with metals. These winds are observed to be highly multiphase, with cool clouds of gas entrained in a fast moving hot wind \citep[see][]{Veilleux2005, Rupke2018}. On small scales in idealized wind tunnel simulations, mixing has been shown to play a key role in cloud evolution. In the adiabatic limit, Kelvin-Helmholtz (KH) and other hydrodynamic instabilities destroy clouds in several characteristic ``cloud crushing times"\citep{Klein1994}. For a cloud with radius $R_{cl}$ in a wind of speed $v_{w}$, and a density contrast $\chi = \frac{\rho_{cl}}{\rho_{w}}$, this timescale is given by

\begin{equation}
t_\mathrm{cc}=\chi^{1/2} \frac{ R_\mathrm{cl}}{v_\mathrm{w}}.
\label{eq:tcc_adiabatic}
\end{equation}

Because the growth rate of KH instabilities in simulations is sensitive to the smallest resolved scale \citep{Chandrasekhar1961}, the rate of mixing will be impacted by the resolution of the simulation. Indeed, previous work has demonstrated that as simulation resolution increases, finer structures appear and the number of cloudlets increases \citep{Cooper2009, Schneider2017}. Despite the important role of mixing, the relationship between cloud evolution and resolution is an under-explored area, particularly in the low resolution limit most relevant in cosmological simulations. In this Note, we explore the effects of numerical resolution on cloud evolution in the simple case of a spherical, adiabatic cloud. We also investigate the role of wind speed by investigating both the supersonic and the less commonly studied subsonic wind regime. 

\begin{figure}[ht!]
\includegraphics[width=\linewidth]{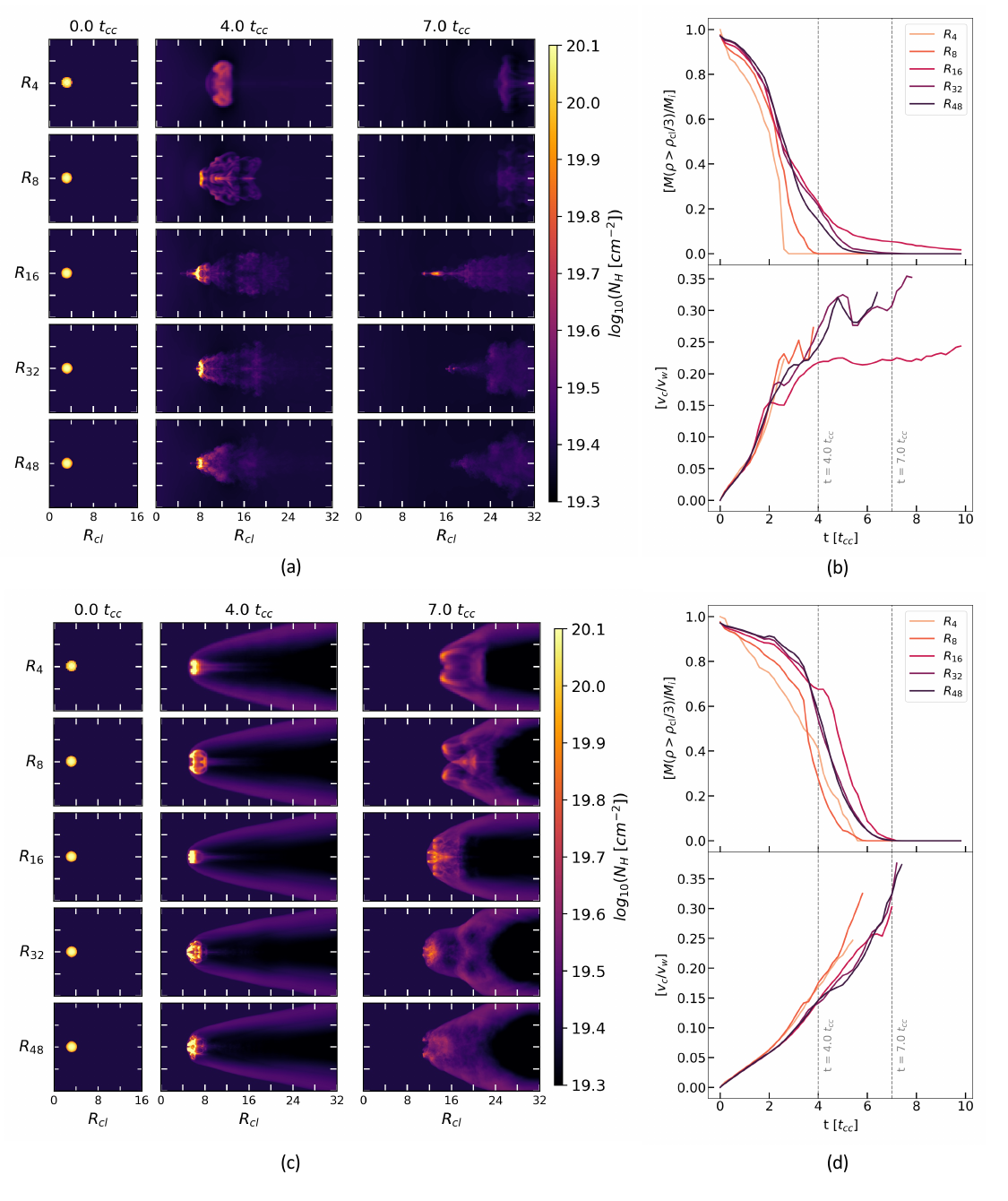}
\caption{The evolution of clouds in two wind speeds: 100 km/s (top) and 1000 km/s (bottom). (a) and (c) show snapshots of column density at resolutions $R_{cl}/\Delta x$ = 4, 8, 16, 32, and 48, as well as at three time-steps: 0, 4, and 7 $t_{cc}$. (b) and (d) show the change in cloud mass over time normalized to the initial cloud mass, and the average cloud velocity normalized by the wind velocity over time.\label{fig:main}}
\end{figure}

\section{Methods and Analysis} \label{sec:methods}

To investigate the effects of resolution on cloud evolution, we ran two sets of simulations at five different resolutions: $R_{cl}/\Delta x$ = 4, 8, 16, 32, and 48 (hereafter $R_{4}$, $R_{8}$, $R_{16}$, $R_{32}$, $R_{48}$). Simulations were run with the Cholla hydrodynamics code \citep{Schneider2015} using PPM reconstruction, an HLLC Riemann solver, and the Van Leer integrator. We employ a standard wind tunnel setup consisting of a long box with a constant wind entering from the left boundary and outflow boundaries on all other sides. All simulations have box dimensions of 32 $R_{cl}$ x 16 $R_{cl}$ x 16 $R_{cl}$. 

In order to explore potential differences in mixing for the subsonic versus supersonic regime, we use wind speeds of 100 km/s (M = 0.852) and 1000 km/s (M = 8.52), respectively. The wind density and temperature are the same for all simulations: $n_{w} = 0.01$ cm$^{-3}$ and $T_{w} = 10^{6}$ K. Clouds are initialized as spheres with a radius of 50 pc and zero initial velocity, and are positioned 3.2 $R_{cl}$ away from the $-x$ boundary (Figure \ref{fig:main}, first column). The cloud-wind density contrast is $\chi$ = $10^2$, and the cloud temperature is $T_{cl} = 10^4$ K . All simulations run for a total of 10 $t_{cc}$.

In the following analysis, we consider cells with a density above $\frac{1}{3}$ of the initial cloud density to be ``cloud" material (Our results are not sensitive to the precise density threshold). We measure cloud mass
\begin{equation}
M_{cl} = \sum \rho_i(> \frac{1}{3}\rho_\mathrm{cl,init}) \mathrm{dV},
\label{eq:M_cl}
\end{equation}
and mass-weighted average cloud velocity, $\bar{v}_\mathrm{x}$,
\begin{equation}
\bar{v}_\mathrm{x}=\frac{ \sum\rho_i(> \frac{1}{3}\rho_\mathrm{cl,init}) v_{x, i}\mathrm{dV}}{M_{cl}},
\label{eq:v_avg}
\end{equation}
every 0.2 $t_{cc}$, where $\rho_i$ is the mass density of the $i$th cell and $\mathrm{dV}$ is the cell volume.

\section{Results \& Discussion} \label{sec:results}

Figure \ref{fig:main} illustrates the difference in cloud evolution at each resolution. Panels (a) and (c) show column density snapshots of the cloud in its initial state, in a partially disrupted state at 4 $t_{cc}$, and late in its evolution at 7 $t_{cc}$. These snapshots clearly show how hydrodynamic instabilities destroy the cloud within several $t_{cc}$ in all cases. However, in both the subsonic and supersonic regimes, the cloud's acceleration and destruction is affected by the resolution of the simulation. Somewhat surprisingly, this trend does not appear to be monotonic. In the subsonic case in particular (panel a) the destruction follows an almost parabolic trend, with a turning point at $R_{16}$. 

In panels (b) and (d) we further investigate this behavior by plotting the normalized cloud mass $M_{cl}/M_{cl_{i}}$ and the normalized average cloud velocity $\bar{v}_{x}/v_{w}$ over time. In (b) we see in the subsonic wind case that the $R_{16}$ cloud survives the longest and is accelerated to significantly lower terminal velocities than the other clouds, reaching a terminal velocity of only ~23\% of the wind speed. In the supersonic wind, (d) shows that while the $R_{16}$ cloud is again the slowest to be completely destroyed, the velocity trend is more monotonic, with the lowest resolutions accelerating the most quickly and the highest resolutions the most slowly. In both subsonic and supersonic regimes (b) and (d) show that the highest speeds are achieved in the highest resolution simulations, with $R_{32}$ reaching ~35\% of the wind speed in both the subsonic and supersonic cases. Results appear to be converging at $R_{32}$ in both cases.

The early-time acceleration at all resolutions is similar, while the late-time behavior of the velocities diverge. This is likely a result of a change in acceleration mechanism throughout the cloud evolution. Ram pressure is the dominant acceleration mechanism in the beginning of the interaction, before mixing becomes efficient and destroys the cloud. While the efficiency of mixing is resolution-dependent, the ram pressure only depends on the surface area of the cloud, which is the same for all simulations. Because the ram pressure of the wind scales as $\rho v^2$, we expect the acceleration due to ram pressure to be 100 times greater in the supersonic models, while the mixing rate scales only linearly with the shear velocity. This may explain why the resolution trend seen in the subsonic versus supersonic simulations is different. 

\begin{acknowledgments}
H.J.L thanks Matthew Abruzzo, Robert Caddy, Alwin Mao, and Orlando Warren for many helpful discussions. This research was supported in part by the University of Pittsburgh Center for Research Computing, RRID:SCR\_022735, through the resources provided. Specifically, this work used the H2P cluster, which is supported by NSF award number OAC-2117681. This material is based upon work supported by the National Aeronautics and Space Administration under Grant No. S000978-NASA issued through the Pennsylvania Space Grant Consortium. 

\end{acknowledgments}

\vspace{3mm}

\software{Cholla \citep{Schneider2015}, \texttt{numpy} \citep{VanDerWalt11}, \texttt{matplotlib} \citep{Hunter07},  \texttt{hdf5} \citep{hdf5} \texttt{seaborn} \citep{Waskom2021}}

\bibliography{res-dependence}{}
\bibliographystyle{aasjournal}

\end{document}